\documentclass[twocolumn,aps, prl, letterpaper]{revtex4}
\usepackage{amsmath}
\usepackage{color}
\usepackage{graphicx}
\usepackage{booktabs}
\usepackage{multirow}
\usepackage{todonotes}
\usepackage{wrapfig}
\usepackage{float}
\usepackage[bookmarks=false,plainpages=false,colorlinks=true,urlcolor=black,citecolor=black,linkcolor=black]{hyperref}
\usepackage{amssymb,bm}
\usepackage{amsthm} 
\usepackage{amsmath}
\usepackage{pifont}

\input{Qcircuit}

\def\ket#1{{\lvert}#1\rangle}
\def\bra#1{\langle #1{\lvert}} 
\def\cross{\ding{55}}
\def\tick{\ding{51}}

\def\be{\begin{equation}}
\def\ee{\end{equation}}

\begin{document}
\title{Experimental verification of quantum computations}

\author{Stefanie Barz$^{1}$, Joseph F. Fitzsimons$^{2,3}$, Elham Kashefi$^{4}$, Philip Walther$^{1}$}
 \affiliation{
$^1$~University of Vienna, Faculty of Physics, Boltzmanngasse 5, 1090 Vienna, Austria,\\
$^2$~Singapore University of Technology and Design, 20 Dover Drive, Singapore~138682,\\
$^3$~Centre for Quantum Technologies, National University of Singapore, Block S15, 3~Science Drive~2, Singapore~117543,\\
$^4$~School of Informatics, University of Edinburgh, 10 Crichton Street, Edinburgh EH8 9AB, UK
}

\vspace{0.5cm}
\begin{abstract}
Quantum computers are expected to offer substantial speedups over their classical counterparts and to solve problems that are intractable for classical computers. Beyond such practical significance, the concept of quantum computation opens up new fundamental questions, among them the issue whether or not quantum computations can be certified by entities that are inherently unable to compute the results themselves. Here we present the first experimental verification of quantum computations. We show, in theory and in experiment, how a verifier with minimal quantum resources can test a significantly more powerful quantum computer. The new verification protocol introduced in this work utilizes the framework of blind quantum computing and is independent of the experimental quantum-computation platform used. In our scheme, the verifier is only required to generate single qubits and transmit them to the quantum computer. We experimentally demonstrate this protocol using four photonic qubits and show how the verifier can test the computer's ability to perform measurement-based quantum computations.
\end{abstract}
\maketitle

The prevalent scientific paradigm of testing physical theories by comparing experimental results with predictions computed on a piece of paper or on a computer assumes that all such predictions are solvable in polynomial time on a classical computer.
In current experiments involving quantum particles, such as fundamental tests of quantum mechanics or small-scale quantum computations and simulations~\cite{Deutsch1985,Deutsch1992, Grover1996, Shor1997, Harrow2009}, following this paradigm is still possible, as the results can be calculated on a classical computer and verified in experiments involving quantum systems. However, there is an entire class of problems---for example, the simulation of complex quantum systems~\cite{Feynman1982}---that are solvable in polynomial time only on a quantum computer~\cite{Watrous2008}.

One of the central conceptual questions in current quantum computing is therefore whether any entity can test the results obtained by a quantum computer, even when that entity is unable to compute these results itself. Or, from a different perspective, can an experimentalist with only classical resources or restricted quantum resources prove that a given device is a quantum computer~\cite{Pappa2012}? Whereas the ultimate answer to such questions is still open, there are several proposals that offer a solution when the verifier is equipped with a range of quantum resources~\cite{Aharonov2008, Broadbent2009, Aharonov2012, Fitzsimons2012, Morimae2012, Reichardt2013}---quantum memory, two entangled quantum computers, or a large number of qubits---which, however, are outside the reach of current technology.

\begin{figure}
\centering
\includegraphics[width=0.45\textwidth]{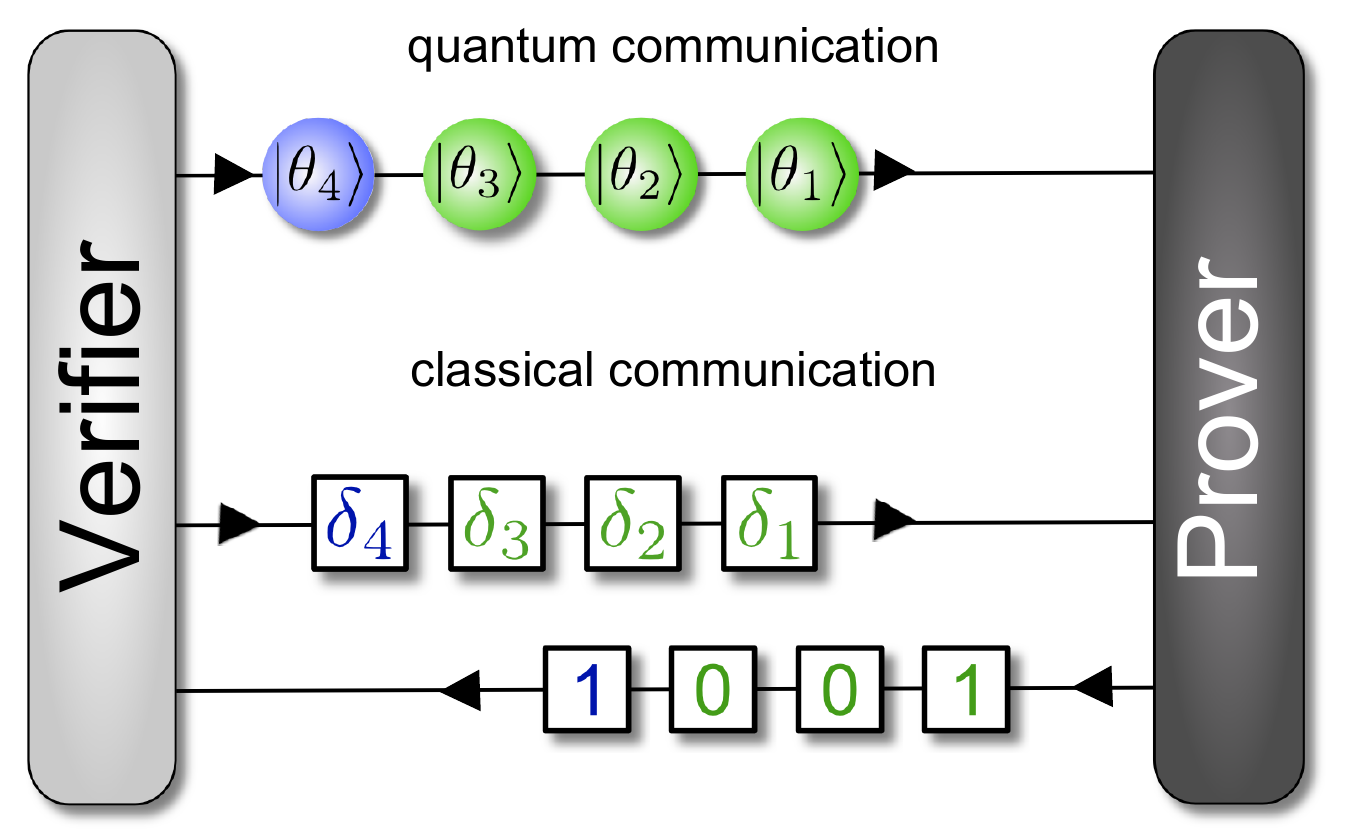}
\caption{\label{figure1}Concept of a quantum prover interactive proof system based on blind quantum computing. The verifier wants to find out if the prover can indeed perform quantum computations. While the question of whether a classical verifier can test a quantum system is still open, it was shown that a verifier who has access to certain quantum resources can verify quantum computations. Here, in the framework of blind quantum computing, the verifier has to be able to generate single qubits and to transmit them to the prover. After the transmission of the qubits, the verifier and the prover exchange two-way classical communication.}
\end{figure}

Here, we demonstrate how to verify a quantum computation on four qubits. Our method is directly applicable to current technology and can be readily extended to more general cases.
We show that only minimal quantum resources (specifically, single qubits) are required to certify a quantum-information processor. Our protocol is independent of the physical system on which it is implemented and it can therefore be applied to any quantum-computing platform.

We have implemented the new protocol on a photonic quantum system and demonstrate the necessary components for verifying a quantum device. We also show how our scheme can be used to verify the generation of the archetype of a quantum-computational resource, quantum entanglement, via a violation of Bell's inequality. In such a verification, the prover remains blind and cannot distinguish the verification procedure from standard quantum-computational tasks such as single- or multi-qubit gates or entire quantum algorithms. To the best of our knowledge, this is the first experiment towards certifying the correctness of a quantum computation.

\section{Interactive proof systems and blind quantum computing}

Our protocol combines interactive proof systems and blind quantum computing~\cite{Aharonov2008,Aharonov2012,Broadbent2009}.
Interactive proof systems were originally invented in the field of computer science, to approach questions in classical complexity theory~\cite{Babai1985, Goldwasser1989}. They have since been extended into the realm of quantum computation~\cite{Watrous2008}. A \textit{quantum prover interactive proof} system addresses the question of whether a prover who has access to quantum-computational resources can convince a classical verifier that he can solve a given problem.
Interactive proof systems can therefore be used to address the fundamental questions posed above, provided the traditional scientific paradigm of ``predicting'' is replaced by ``verifying''.

In our protocol, the framework of the interactive proof system is given by blind quantum computing (Fig.~\ref{figure1}).
In this framework, a verifier (or client) with limited quantum computational resources can delegate a quantum computation to a prover (or server) with the full power of quantum computing such that all data and the whole computation remain private~\cite{Broadbent2009,Morimae2012a}.
More specifically, the verifier prepares single qubits in the state
\begin{equation}
	\ket{\theta_j}=\frac{1}{\sqrt{2}}\left(\ket{0}+e^{i\theta_j}\ket{1}\right)
\end{equation}
with $\theta_j\in\left\{0,\pi/4, ..., 7\pi/4\right\}$ chosen uniformly at random and only known to the verifier. The qubits are then transmitted to the prover who entangles them to create a blind cluster state~\cite{Barz2012}. The actual computation is measurement-based~\cite{Raussendorf2001, Raussendorf2003}.
The verifier calculates for each blind qubit measurement instructions according to
\begin{equation}
	\delta_j=\theta_j+\phi_j+\pi r_j
\end{equation}
where $\theta_j$ is the blind phase of the qubit, $\phi_j$ is the rotation that the verifier wants to perform (including any Pauli corrections), and $r_j\in\{0,1\}$ is a randomly chosen value to hide the measurement outcome.
The prover performs measurements in the basis
\begin{equation}
	\ket{\pm_{\delta_j}}=\frac{1}{\sqrt{2}}\left(\ket{0}\pm e^{i\delta_j}\ket{1}\right)
\end{equation}
 and delivers the results to the verifier.
 Without the knowledge of the underlying rotation and the random phase, the prover cannot find out anything about the actual rotation $\phi_j$---thus the computation remains blind. The verifier, in contrast, knows the initial rotation and is able to interpret the results. Blind quantum computing therefore provides a powerful tool to delegate computations and to access the resources of powerful quantum computers without divulging the content of the computation. In the following, we show how this concept can be applied to verify quantum computations.

\begin{figure}
\centering
\includegraphics[width=0.45\textwidth]{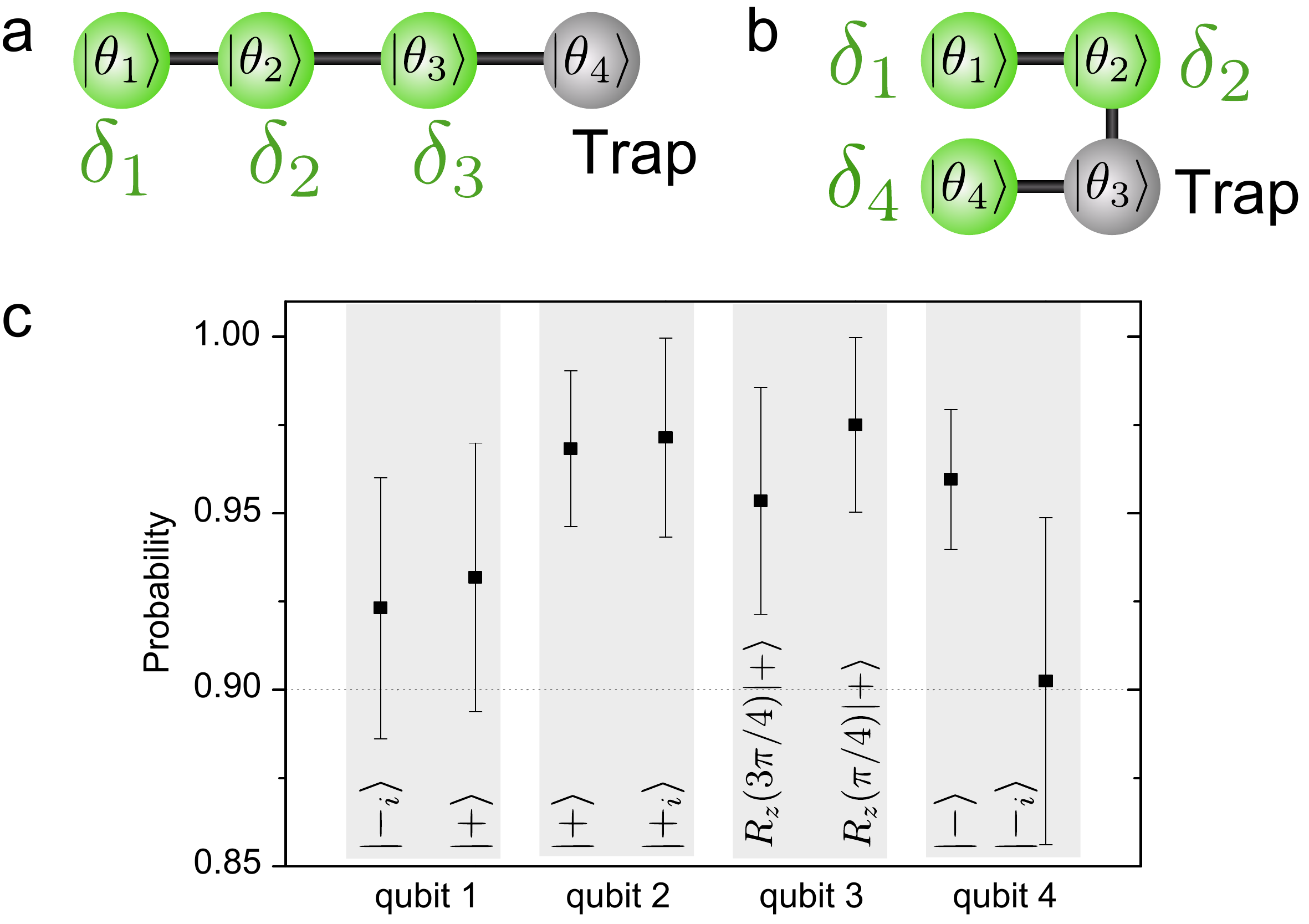}
\caption{\label{figure2}Measurement verification. a,b) A blind linear cluster state and a blind rotated horseshoe cluster state which can be used for the preparation of trap qubits. c) Experimental results of the measurement verification. We prepare two different trap state on each qubit~1-4 and show the probability of obtaining the correct outcome when measuring those qubits.}
\end{figure}
\section{Verification of a quantum computation}
In the framework of blind quantum computing, in order to test a quantum computation, the correctness of the measurements performed by the server has to be verified. Here we use a verification procedure that is based on the creation of trap qubits~\cite{Fitzsimons2012}.
Trap qubits are blindly prepared in a well-defined state, which is only known to the verifier, and are isolated from the actual computation. The measurement angle of these trap qubits is chosen such that the measurement result is predetermined by the verifier, and hence any cheating strategy used by the server that alters these measurement outcomes will be detected. By randomly choosing the locations of the trap qubits it is then possible to bound the probability that the server can cheat while remaining undetected.

In our setting, we implement the preparation of the trap qubits through a measurement-based computation on the non-trap qubits.
The verifier chooses measurement settings on the cluster state such that any of the qubits could become a trap qubit,  prepared in a random state $\ket{\theta_j}$. If the trap qubit is then measured in the basis $\ket{\theta_j}$, the outcome will always be known to the verifier (see Fig.~\ref{figure2}).  Our measurement-based creation of the trap qubits means that we verify a correlation between a subset of measurements, rather than a single measurement outcome. We therefore have to be careful to ensure that the correctness of these correlations for all trap measurements does imply the correctness of a given computational run.
In our demonstration using a four-qubit system, only one error remains undetected (see Appendix). 
However, this particular error cannot alter the result of the measurement of Bell's quantity presented in this paper.

\begin{figure}
\centering
\includegraphics[width=0.45\textwidth]{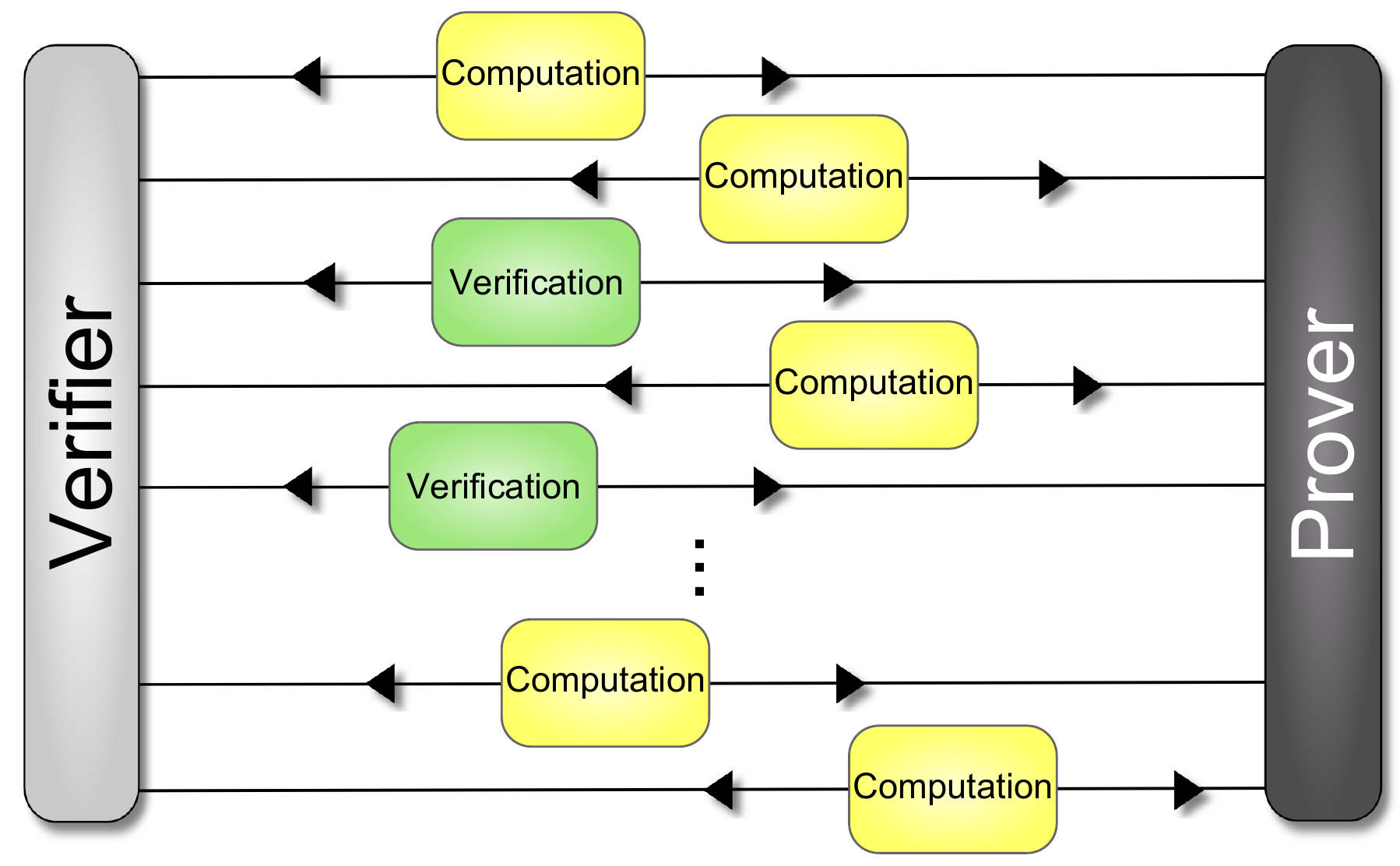}
\caption{\label{figure3}Schematic of a quantum computation with verification sub-routines.}
\end{figure}

This verification procedure can be used to verify that the quantum computation was performed correctly.
Therefore, we consider multiple runs of the protocol, where the verifier randomly choses to run an actual computation or a verification test (see Fig.~\ref{figure3}).
This use of multiple runs of the blind-computation protocol lets us make optimal use of the qubits available in our system. Moreover, the server cannot distinguish between an actual computation run or a trap run. Hence, as discussed in details in the appendix, this procedure can be used to verify not only the correctness of the measurement outcomes but also of the entire quantum computation. When trap computation and target computation are randomly interspersed, then the probability that the quantum computer produces the correct result for the verification runs but a wrong result for the computation runs is bounded by a value depending on three parameters: the number of computation runs, the number of trap runs, and on the total number of qubits in the system (see appendix).

\section{Entanglement verification}

\begin{figure}
\centering
\includegraphics[width=0.45\textwidth]{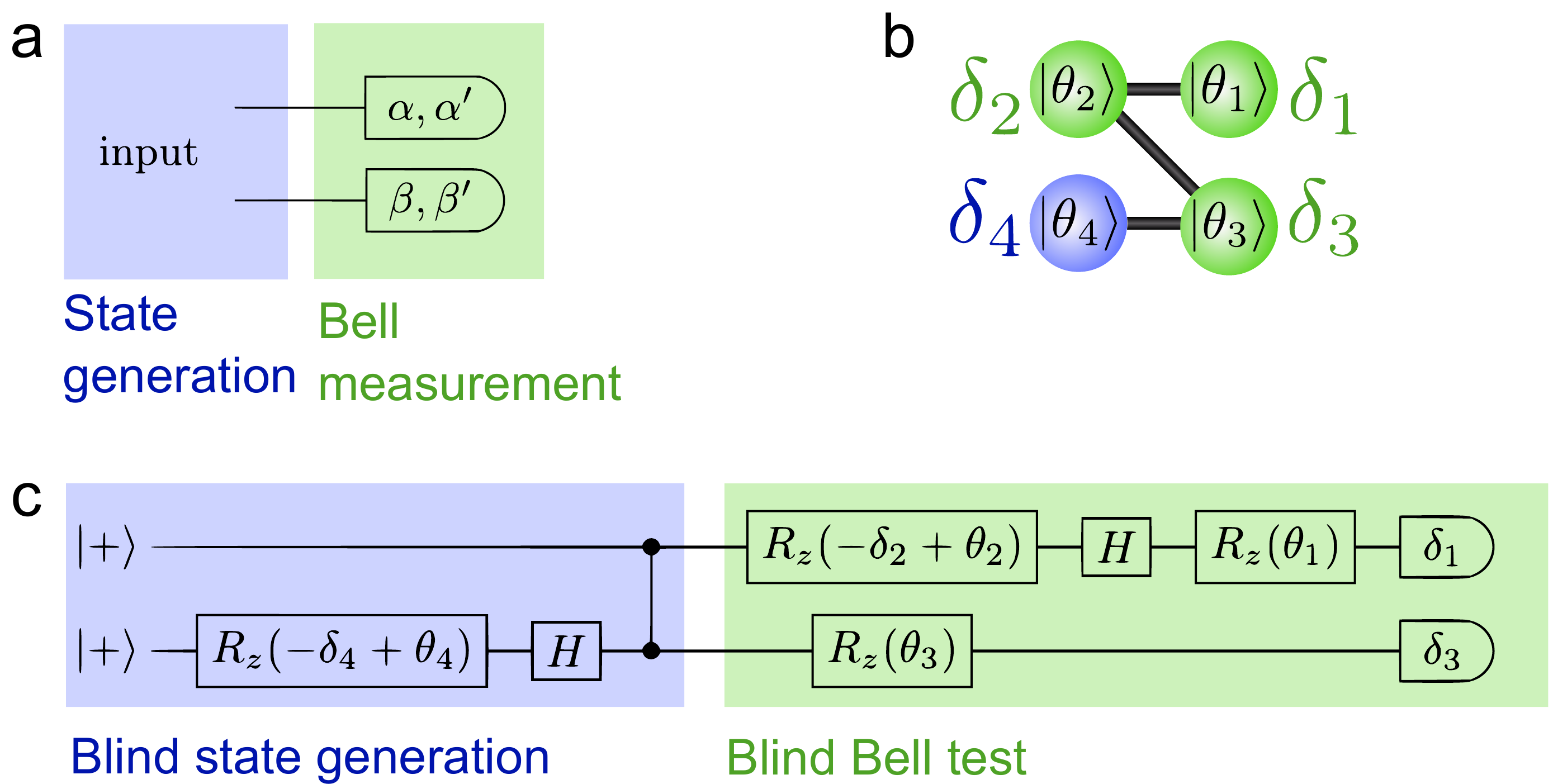}
\caption{\label{figure4}A blind Bell test for the verification of quantum resources.
a) Conventional scheme for a Bell test, where first an (entangled) state is created and then Bell measurements are performed. b) The blind zigzag cluster state, and c) its corresponding circuit.
If the rotation in the lower wire, $-\delta_4+\theta_4$, is chosen equal to zero (or $\pi$), the input state in the lower wire will be equal to $\ket{0}_b$ ($\ket{1}_b$); otherwise, if the rotation is chosen equal to $\pm\pi/2$ the input will be $\ket{\pm_i}_b$.
The edge between qubits 2 and 3 performs a \mbox{CPhase} gate on the two qubits, which results  in an entangled state in the former case, and in an unentangled state in the latter. The values of $\delta_1$,  $\delta_2$, and $\delta_3$, as well as the phases $\theta_1$,  $\theta_2$, and $\theta_3$ determine the Bell measurement settings.
}
\end{figure}

Once the measurement outcomes are verified, we can proceed to use the system to probe the prover's entangling capabilities and its ability to create cluster states.
Quantum correlations are typically confirmed by well-established tests of Bell's inequality~\cite{Bell1964} (Fig.~\ref{figure4}). In order to do so, combinations of specific measurement settings $\alpha, \alpha'$ and $\beta, \beta'$ are performed on the first ($a$) and on the second qubit ($b$), respectively, after generating an entangled state $\ket{\Psi}_{a,b}$.
The settings are chosen such that a maximal violation of the Bell inequality of the Clauser-Horne-Shimony-Holt (CHSH) type~\cite{Clauser1969} is obtained for an entangled state:

\begin{equation}
	S=|E(\alpha, \beta)-E(\alpha, \beta')|+|E(\alpha', \beta)+E(\alpha', \beta')|\leq2
\end{equation}
The correlation coefficients $E(\cdot, \cdot)$ are defined by the coincidence counts when measuring qubit $a$ in the basis $\alpha$ and qubit $b$ in the basis $\beta$ (for details see appendix).

In order to make the Bell test blind, we hide the generation of the entangled state as well as the Bell measurement settings.
For this, we base our implementation on a blind zigzag cluster state with four qubits $\ket{\theta_j}$, which is shown in Fig.~\ref{figure4}b.
Single-qubit measurements on the blind zigzag cluster state realize a quantum circuit that offers exactly the degrees of freedom that are necessary for our blind Bell test.
First, using this type of cluster, the verifier has the possibility to blindly switch between
entangled or separable input states by choosing $\delta_4$ and $\theta_4$ accordingly.
Second, the standard  measurement settings for a Bell test are hidden as they are determined by the phases of the blind qubits $\ket{\theta_1}$, $\ket{\theta_2}$, and $\ket{\theta_3}$ and their respective measurement settings of $\delta_1$, $\delta_2$, and $\delta_3$ (see appendix for details).

As a result, the state generation as well as the Bell-state measurements are encoded in the phase of blind qubits as well as in the measurement instructions, which remain unknown to the prover at any time. The choice of the cluster-state configuration also remains hidden from the prover. This is a particular advantage of our probabilistic implementation of blind quantum computing, where all qubits are measured.

\section{Experiment}

 We use the particular advantages offered by photons to realize a quantum network that can communicate and process quantum information~\cite{Knill2001} within the same physical system~\cite{Barz2012}. In our experiment, the blind cluster states are generated from photon pairs entangled in polarization and mode, which originate from spontaneous parametric down-conversion~\cite{Kwiat1995}. Our setup and the methods used are explained in detail in ref. 10; in the present experiment, we generate blind cluster states for various settings of $\theta_j$ and use them to implement exemplary runs of trap computations as well as the Bell-test runs---the necessary building blocks of a verified test of Bell's quantity.

For the demonstration of the measurement verification, we use blind linear cluster states and blind rotated horse-shoe cluster state to prepare traps as shown in Fig.~\ref{figure2}a and~\ref{figure2}b.
By choosing the blind phases $\theta_j$ and measurement settings $\delta_j$ as given in the appendix, we prepare the traps:
\begin{eqnarray}
	\ket{\mbox{trap}_1}&=&\ket{-_i} \mbox{ and } \ket{+}\\
	\ket{\mbox{trap}_2}&=&\ket{+} \mbox{ and } \ket{+_i}\\
	\ket{\mbox{trap}_3}&=&R_z(3\pi/4)\ket{+} \mbox{ and } R_z(\pi/4)\ket{+}\\
	\ket{\mbox{trap}_4}&=&\ket{-} \mbox{ and } \ket{-_i}
	\label{eqn:traps}
\end{eqnarray}
on qubits 1, 2, 3, and 4, respectively ($\ket{\pm}=(\ket{0}\pm \ket{1})/\sqrt{2}$, $\ket{\pm_i}=(\ket{0}\pm i \ket{1})/\sqrt{2}$).
The probabilities to find the correct outcome are well above $90\%$, as shown in Fig.~\ref{figure2}c.

For a verification of quantum entanglement, we choose combinations of $\theta_4$ and $\delta_4$ to create the entangled state:
\begin{equation}\label{entangledstate}
	\ket{\Psi}_{a,b}=\frac{1}{\sqrt{2}}(\ket{+}_a\ket{0}_b-i \ket{-}_a\ket{1}_b)
\end{equation}
and to blindly implement the Bell measurements settings we choose:
\begin{equation}\label{eq:Bellangles}
\alpha=\pi/2,\,\ \alpha'=\sigma_z,\,\ \beta=-3\pi/4,\,\ \beta'=-\pi/4
\end{equation}
where the bases $\alpha$, $\beta$, and $\beta'$ are defined as $\{(\ket{0}\pm e^{i\alpha}\ket{1})/\sqrt{2}\}$ etc., and $\alpha'$ is a measurement in the basis $\{\ket{0},\ket{1}\}$.
To obtain the measurement settings given in Eq.~(\ref{eq:Bellangles}), we choose combinations of $\ket{\theta_j}$ and $\delta_j$ as given in detail in the appendix.
From the measured coincidence count rates, we calculate the correlation coefficients to be:
\begin{eqnarray}
	E(\alpha, \beta)&=& -0.540\pm0.084\\
	E(\alpha, \beta')&=& 0.634\pm0.086\\
	E(\alpha', \beta)&=& -0.646\pm0.067\\
	E(\alpha', \beta')&=& -0.678\pm0.079.
\end{eqnarray}
Those coefficients lead to an S parameter of
\begin{equation}
	S=2.498\pm0.158,
\end{equation}
 which violates the classical bound ($\left|S\right|=2$) by more than 3 standard deviations.

The combination of the verification procedure and this violation of Bell's inequality suffices to unambiguously verify the prover's ability to perform entangling gates between qubits and thus to create cluster states. In our experiment, we implement a subset of all possible blind states. The states of qubits~1 and~4 are fixed to $\ket{+}$, whereas the states of qubits~2 and~3 are fully blind~\cite{Barz2012}.
The whole verification procedure remains blind, however, if we assume that the prover has no \textit{a priori} knowledge of our choice of states and measurements.

The violation of Bell's inequalities is impossible classically, not for the reason of high complexity, but rather on the basis of physical principles. In our implementation, we assume the correctness of quantum mechanics for the verification of the measurement outcomes. Without this assumption, a full demonstration would require the two entangled photons to be sent to two distant laboratories, where only at the very last step of the computation the verifier gives the measurement instructions to the prover. In this way, no classical computer could mimic, even in principle the output of the computation \textit{a priori}, while the verification procedure would still have a positive outcome.

\section{Conclusion}
Future large-scale quantum computers and quantum simulators~\cite{Weitenberg2011, Islam2011, Monz2011, Britton2012} will require the verification of their experimental results~\cite{Leibfried2010}. Due to the superior computational capacity of quantum systems the results cannot simply be calculated and checked on a classical device. The development of new methods for the verification of quantum computations is therefore a crucial task.
Here, we have developed a new general method for verifying quantum computations that can be readily applied to current small-scale quantum computers.
We have shown, in theory and in experiment, how a verifier can test whether the quantum computer is quantum and even whether it computes correctly.

Finally, how verification mechanisms provide insights into questions of computational complexity
and into the foundations of quantum physics it is a topic of active current research. To date, the limit of high computational complexity is mostly unexplored and  it is not impossible that quantum mechanics breaks down at some scale of complexity~\cite{Aaronson2004}.

Verification methods, such as those reported here in are not only important as a mechanism to certify quantum computers, but also provides an entirely novel toolbox for addressing fundamental questions in quantum physics and computer science.



{\bf Acknowledgment} The authors are grateful to  S. Aaronson, D. Aharonov, C. Brukner, A. Zeilinger, and F. Verstraete for discussions.
S.B. and P.W. acknowledge support from the European Commission, Q-ESSENCE (No. 248095), QUILMI (No. 295293) and the ERA-Net CHISTERA project QUASAR, the John Templeton Foundation, the Vienna Center for Quantum Science and Technology (VCQ), the Austrian Nano-initiative NAP Platon, the Austrian Science Fund (FWF) through the SFB FoQuS (No. F4006-N16), START (No. Y585- N20) and the doctoral programme CoQuS, the Vienna Science and Technology Fund (WWTF) under grant ICT12-041, and the Air Force Office of Scientific Research, Air Force Material Command, United States Air Force, under grant number FA8655-11-1-3004.
J.F. acknowledges support from the National Research Foundation and the Ministry of Education, Singapore. E.K. acknowledges support from UK Engineering and Physical Sciences Research Council (EP/E059600/1).

\newpage
\begin{widetext}
\section{Appendix}

\section{Verification of a quantum computation}

In order to verify a blind quantum computation, it is necessary to ensure that the probability of an undetected error being introduced to the computation is bounded. One way to do this is to introduce trap qubits into the computation as in~\cite{Fitzsimons2012}. To prove that this does in fact guarantee that any error is either detected or corrected except with bounded probability, we must consider the most general possible cheating strategy for the prover. Thus we must consider the effect of an arbitrary deviation at each step of the protocol. In this section we present a simplified version of the proof in~\cite{Fitzsimons2012} adapted to our 4-qubit protocol with classical inputs and outputs, and then show how it can be adapted to work with traps prepared by measurement-based computation.
 
\subsection{Individual trap qubits}
We assume the most general scenario.
The prover obtains the quantum states $\ket{\theta_i}$ and the states $\ket{\delta_i}$ which encode the classical angles $\delta_i$. 
Further, the prover has access to a private quantum memory $\ket{\text{Prover}}$, where the prover could store quantum information allowing him to perform the most general attacks:

\[
\Qcircuit @C=0.5em @R=0.5em @!R {
\lstick{\ket{\theta_1}} & \multigate{4}{B_1} &  \multigate{5}{B_2} & \meter& b_1\\
\lstick{\ket{\theta_2}} & \ghost{B_1} &  \ghost{B_2} &\multigate{5}{B_3}  & \meter& b_2\\
\lstick{\ket{\theta_3}} & \ghost{B_1} & \ghost{B_2} &  \ghost{B_2} &  \multigate{5}{B_4}  & \meter & b_3\\
\lstick{\ket{\theta_4}} & \ghost{B_1} &  \ghost{B_2}  & \ghost{B_2} &  \ghost{B_2}  & \multigate{5}{B_5} & \meter & & b_4 \\
\lstick{\ket{\text{Prover}}} & \ghost{B_1} & \ghost{B_2} &  \ghost{B_2} &  \ghost{B_2}  & \ghost{B_2}  & \qw & \qw & \qw\\
\lstick{\ket{\delta_1}} &  \qw & \ghost{B_2}  & \ghost{B_2} &  \ghost{B_2}  & \ghost{B_2} & \qw & \qw &\qw \\
& \lstick{\ket{\delta_2}} & \qw &  \ghost{B_2} &  \ghost{B_2}  & \ghost{B_2} & \qw & \qw &\qw\\
&  & \lstick{\ket{\delta_3}} & \qw &  \ghost{B_2}  & \ghost{B_2} & \qw &\qw &\qw\\
 &  &  &  \lstick{\ket{\delta_4}}& \qw   & \ghost{B_2} &  \qw &\qw &\qw
}\]

Here,  $\{B_i\}$ are the individual operations performed by the prover and $b_i$ is the outcome of a measurement always performed in the basis $\{\ket{0}, \ket{1}\}$.
Without loss of generality we can assume that the measurement occurs immediately prior to transmission of each bit to the verifier, as shown above. Mathematically, the individual operations performed by the prover, $\{B_i\}$ can be combined into a single operation $B$, resulting in the quantum circuit shown below.
\[
\Qcircuit @C=1em @R=0.5em @!R {
\lstick{\ket{\theta_1}} & \multigate{8}{~~~B~~~} & \meter& b_1\\
\lstick{\ket{\theta_2}} & \ghost{~~~B~~~}  & \meter& b_2\\
\lstick{\ket{\theta_3}} & \ghost{~~~B~~~} & \meter & b_3\\
\lstick{\ket{\theta_4}} & \ghost{~~~B~~~} &  \meter & b_4 \\
\lstick{\ket{\text{Prover}}} & \ghost{~~~B~~~} & \qw & \qw\\
\lstick{\ket{\delta_1}} & \ghost{~~~B~~~}  & \qw & \qw\\
\lstick{\ket{\delta_2}} & \ghost{~~~B~~~} & \qw & \qw\\
\lstick{\ket{\delta_3}} & \ghost{~~~B~~~}  & \qw & \qw\\
\lstick{\ket{\delta_4}}& \ghost{~~~B~~~} & \qw &\qw
}\]

By defining $B' = B P^\dagger$, where $P$ corresponds to the unitary implementing the protocol for an honest prover, the circuit corresponding to the protocol can be rewritten as the ideal protocol followed by some deviation.
\[
\Qcircuit @C=1em @R=0.5em @!R {
\lstick{\ket{\theta_1}} & \multigate{8}{~~~P~~~}  & \multigate{8}{~~~B'~~~} & \meter& b_1\\
\lstick{\ket{\theta_2}}  & \ghost{~~~P~~~} & \ghost{~~~B'~~~}  & \meter& b_2\\
\lstick{\ket{\theta_3}} & \ghost{~~~P~~~} & \ghost{~~~B'~~~} & \meter & b_3\\
\lstick{\ket{\theta_4}}& \ghost{~~~P~~~} & \ghost{~~~B'~~~} &  \meter & b_4 \\
\lstick{\ket{\text{Prover}}}& \ghost{~~~P~~~} & \ghost{~~~B'~~~} &\qw & \qw\\
\lstick{\ket{\delta_1}} & \ghost{~~~P~~~} & \ghost{~~~B'~~~}  & \qw & \qw\\
\lstick{\ket{\delta_2}} & \ghost{~~~P~~~} & \ghost{~~~B'~~~} & \qw & \qw\\
\lstick{\ket{\delta_3}} & \ghost{~~~P~~~}  & \ghost{~~~B'~~~}  & \qw & \qw\\
\lstick{\ket{\delta_4}} & \ghost{~~~P~~~} & \ghost{~~~B'~~~} & \qw &\qw
}\]

The verifier's output only corresponds to the measurement outputs received from the prover, and so the prover's effective deviation operator can be reduced to a super-operator, dependent on the specific values of $\{\delta_i\}$ used in that run of the protocol, acting only on these qubits.

\[
\Qcircuit @C=1em @R=0.5em @!R {
\lstick{\ket{\theta_1}} & \multigate{3}{~~~P_\delta~~~}  & \multigate{3}{~~~B'_\delta~~~} & \meter& b_1\\
\lstick{\ket{\theta_2}}  & \ghost{~~~P_\delta~~~} & \ghost{~~~B'_\delta~~~}  & \meter& b_2\\
\lstick{\ket{\theta_3}} & \ghost{~~~P_\delta~~~} & \ghost{~~~B'_\delta~~~} & \meter & b_3\\
\lstick{\ket{\theta_4}}& \ghost{~~~P_\delta~~~} & \ghost{~~~B'_\delta~~~} &  \meter & b_4 
}\]

As the left hand part of the above circuit implements the ideal protocol, $B'_\delta$ contains any error introduced by the prover.

Note that the output of the ideal protocol $b= \{b_i\}$ is the output of the verifier's chosen computation, $m=\{m_i\}$, bitwise xored with a random bitstring, $r = \{r_i\}$, known only to the verifier:
\begin{equation*}
b=m\oplus r.
\end{equation*}

In the following, we encode the classical information in a quantum system:
\begin{eqnarray*}
b&\rightarrow&\ket{b},\\
m&\rightarrow&\ket{m},\\
r&\rightarrow&\ket{r}, \mbox{and}\\
(m\oplus r) &\rightarrow& \ket{m\oplus r}, 
\end{eqnarray*}
thus n classical bits are encoded in n qubits.
 
 Therefore, for a fixed computation chosen by the verifier with outcome $m$, on $n$ qubits, the probability of an error occurring (averaging over all possible choices for the random bitstring $r$) is given by 

\begin{eqnarray*}
\epsilon &=& \frac{1}{2^n}\sum_{r\in\{0,1\}^n}\mbox{Tr}\left[\,\left(\mathbb{I}_{2^n}-\ket{m+r}\bra{m+r}\right)\,B'_\delta\,\left(\ket{m+r}\bra{m+r}\right)\right],\\
&=&\frac{1}{2^n}
\sum_{b\in\{0,1\}^n}\mbox{Tr}\left[\,\left(\mathbb{I}_{2^n}-\ket{b}\bra{b}\right)\,B'_\delta\,\left(\ket{b}\bra{b}\right)\right],
\end{eqnarray*}
for a fixed computation.
Here, $n$ is the number of qubits involved in the protocol and $\mathbb{I}_{2^n}$ is the $2^n$-dimensional density matrix.

Note, however, that for a trap located on any measured qubit $i$ for which the expected measurement outcome is $r_i$, the probability of the trap registering an error is:
\begin{equation*}
	t_{i,r_i} = \mbox{Tr}\left[\left(\mathbb{I}_{2^n} -  \mathbb{I}_{2^{i-1}}\otimes \ket{r_i}\bra{r_i}\otimes \mathbb{I}_{2^{n-i}}\right) B'_\delta \left(\frac{\mathbb{I}_{2^{i-1}}}{2^{i-1}}\otimes \ket{r_i}\bra{r_i}\otimes \frac{\mathbb{I}_{2^{n-i}}}{2^{n-i}}\right)\right],
\end{equation*}
where $\ket{r_i}$ a 2-dimensional quantum state encoding the classical value $r_i$.
Here, we use the fact that for a measurement of a trap qubit we expect $m_i=0$, and thus $r_i=b_i$.

Averaging over all possible choices of $i$ and $r_i$, this yields an average probability of detection of
\begin{eqnarray*}
\langle t \rangle 
&=& 
\sum_{i=1}^n \frac{1}{n} \sum_{b\in\{0,1\}^{n}} \frac{1}{2^n} \mbox{Tr}\left[\left(\mathbb{I}_{2^n}- \mathbb{I}_{2^{i-1}}\otimes\ket{b_i}\bra{b_i}\otimes \mathbb{I}_{2^{n-i}}\right) B'_\delta \left(\ket{b} \bra{b} \right)\right]\\
&=&  \frac{1}{n\, 2^n} \sum_{b\in\{0,1\}^{n}} \mbox{Tr}\left[\left(\sum_{i=1}^n\left(\mathbb{I}_{2^n} - \mathbb{I}_{2^{i-1}}\otimes\ket{b_i}\bra{b_i}\otimes \mathbb{I}_{2^{n-i}}\right)\right) B'_\delta \left(\ket{b} \bra{b} \right)\right],
\end{eqnarray*}
where $\ket{b_i}_i$ a 2-dimensional quantum state encoding the classical value $b_i$.

As the $B'_\delta \left(\ket{b} \bra{b} \right)$ is positive semi-definite, and
\[\left(\mathbb{I}_{2^n} - \ket{b}\bra{b}\right) \preceq \sum_{i=1}^n \left(\mathbb{I}_{2^n} - \ket{b_i}_i\bra{b_i}_i\right),\]
then
\[
\langle t \rangle \geq \frac{2^{-n}}{n} \sum_{b\in\{0,1\}^{n}} \mbox{Tr}\left[\left(\mathbb{I}_{2^n} - \ket{b}\bra{b}\right) B'_\delta \left(\ket{b} \bra{b} \right)\right]\\
\]
where $b_i$ is the $i$th bit of $b$. Thus, by substituting in $\epsilon$ into the above equation and rearranging, we obtain $\epsilon \leq n \langle t \rangle$.

\subsection{Traps prepared by MBQC}

Contrary to the protocol described in~\cite{Fitzsimons2012}, in the current experiment we rely on measurement-based computation to prepare isolated trap qubits (instead of preparing them directly). For example to prepare qubit 4 as a trap qubit, we choose a blind linear cluster state which implements the following computation:
\begin{equation*}	
	\ket{\mbox{trap}_4}=R_z\left(\theta_4\right)HR_z\left(m_3 \pi\right)H R_z\left(\frac{\pi}{2} + m_2 \pi\right)H R_z\left(\frac{\pi}{2} + m_1 \pi\right)\ket{+}.
\end{equation*}
The output state $\ket{\mbox{trap}_4}$ then depends on the outcomes of the measurement of qubit 1, 2 and 3 which are blind to the prover.

The general measurement patterns which can achieve such isolated trap qubits, together with the corresponding state of the trap qubit prepared are shown in Table \ref{dummyprep}.

\begin{table}[h!]
	\centering
\begin{tabular}{|c|c|c|c|c|c|}
\hline
Trap qubit & \multicolumn{4}{|c|}{Measurements} & Trap state \\
 & 1 & 2 & 3 & 4 & \\
\hline
1 &  & $\sigma$ & $Y$ & $Y$ & $\ket{+_{(m_3 \oplus m_4) \pi}}$\\
2 & $Y$ & & $X$ & $Y$ & $\ket{+_{(m_1 \oplus m_3 \oplus m_4) \pi}}$\\
3 & $Y$ & $X$ & & $Y$ & $\ket{+_{(m_1 \oplus m_2 \oplus m_4) \pi}}$\\
4 & $Y$ & $Y$ & $\sigma$ & & $\ket{+_{(m_1 \oplus m_2) \pi}}$\\
\hline
\end{tabular}
\caption{Measurement choices for non-trap qubits which prepare isolated trap qubits at each location. Note that if the choice of measurement operator for a given qubit does not affect the outcome then the measurement has been denoted by $\sigma$.\label{dummyprep}
}
\end{table}

In order to determine the affect of a cheating prover, it is convenient to note that each of these trap measurements can also be interpreted as a stabilizer measurement of the underlying cluster state, as shown in Table \ref{tab:stabs}. Table \ref{trapqubits} gives the cluster state measurement angles $\phi$ and sample corresponding pairs of blind state preparation and measurement angles $(\theta,\delta)$, together with the classical computation performed in each case to verify the outcome of the trap measurement.
\begin{table}
	\centering
\begin{tabular}{|c|c|}
\hline
Trap qubit & Stabilizer \\
\hline
1 & $X\otimes \mathbb{I} \otimes Y \otimes Y$\\
2 & $Y\otimes X \otimes X \otimes Y$\\
3 & $Y\otimes X \otimes X \otimes Y$ \\
4 & $Y\otimes Y \otimes \mathbb{I} \otimes X$\\
\hline
\end{tabular}
\caption{Index of trap qubit and corresponding stabilizer measurement.\label{tab:stabs}}
\end{table}

\begin{table}[h!]
	\centering
\begin{tabular}{|c||c|c|c|c||c|c|c|c||c|}
\hline
Trap qubit & \multicolumn{4}{|c||}{$\phi$} & \multicolumn{4}{|c||}{Sample $(\theta_i,\delta_i)$} & Trap outcome \\
 & 1 & 2 & 3 & 4 & 1 & 2 & 3 & 4 &\\
\hline
&&&&&&&&&\\

1 & $0$ & $0$ & $\frac{\pi}{2}$ & $\frac{\pi}{2}$ & $(0,0)$ & $\left(\frac{\pi}{2},-\frac{\pi}{2}\right)$ & $\left(\frac{3\pi}{4},\frac{5\pi}{4}\right)$ & $\left(0,\frac{\pi}{2}\right)$ &$m_1 \oplus m_3 \oplus m_4$\\&&&&&&&&&\\
2 & $\frac{\pi}{2}$ & $0$ & $0$ & $\frac{\pi}{2}$ & $\left(0,\frac{\pi}{2}\right)$ & $\left(\frac{\pi}{2},\frac{\pi}{2}\right)$ & $\left(\frac{3\pi}{4},\frac{7\pi}{4}\right)$ & $\left(0,\frac{\pi}{2}\right)$ &$m_1 \oplus m_2 \oplus m_3 \oplus m_4$\\&&&&&&&&&\\
3 & $\frac{\pi}{2}$ & $0$ & $0$ & $\frac{\pi}{2}$ & $\left(0,\frac{\pi}{2}\right)$ & $\left(\frac{\pi}{2},\frac{\pi}{2}\right)$ & $\left(\frac{3\pi}{4},\frac{7\pi}{4}\right)$ & $\left(0,\frac{\pi}{2}\right)$ &$m_1 \oplus m_2 \oplus m_3 \oplus m_4$\\&&&&&&&&&\\
4 & $\frac{\pi}{2}$ & $\frac{\pi}{2}$ & $0$ & $0$ & $\left(0,-\frac{\pi}{2}\right)$ & $\left(\frac{\pi}{2},0\right)$ & $(\pi,0)$ & (0,0) &$m_1 \oplus m_2 \oplus m_4$\\&&&&&&&&&\\
\hline
\end{tabular}
\caption{Measurement angle for each trap setting together with sample $\delta$ and $\theta$. The trap outcome should be consistent with the equation in the right most column in the above table, and any discrepancy represents the detection of an error. \label{trapqubits}}
\end{table}

As in the previous section, a general deviation by the prover can be modeled by the quantum circuit below.
\[
\Qcircuit @C=1em @R=0.5em @!R {
\lstick{\ket{\psi_1}} & \multigate{3}{~~~P_\delta~~~}  & \multigate{3}{~~~B'_\delta~~~} & \meter& b_1\\
\lstick{\ket{\psi_2}}  & \ghost{~~~P_\delta~~~} & \ghost{~~~B'_\delta~~~}  & \meter& b_2\\
\lstick{\ket{\psi_3}} & \ghost{~~~P_\delta~~~} & \ghost{~~~B'_\delta~~~} & \meter & b_3\\
\lstick{\ket{\psi_4}}& \ghost{~~~P_\delta~~~} & \ghost{~~~B'_\delta~~~} &  \meter & b_4 
}\]
Equivalently this can be rewritten as 
\[
\Qcircuit @C=1em @R=0.5em @!R {
\lstick{\ket{+}} & \multigate{3}{~~~G~~~}   & \gate{R_Z(\phi_1)} & \gate{H} & \multigate{3}{~~~B''_\delta~~~}  & \meter& b_1\\
\lstick{\ket{+}} & \ghost{~~~G~~~} & \gate{R_Z(\phi_2)} & \gate{H}  & \ghost{~~~B''_\delta~~~} & \meter& b_2\\
\lstick{\ket{+}} & \ghost{~~~G~~~} & \gate{R_Z(\phi_3)} & \gate{H} &  \ghost{~~~B''_\delta~~~} & \meter & b_3\\
\lstick{\ket{+}} & \ghost{~~~G~~~} & \gate{R_Z(\phi_4)} & \gate{H} &  \ghost{~~~B''_\delta~~~} & \meter & b_4 
}\]
Here, the left hand portion of the circuit performs the unitary part of the ideal measurement-based computation, immediately prior to measurement in the computational basis, with $G$ representing the entangling gate which generates the cluster state from separable qubits via a series of controlled-$Z$ operations. The deviation operator $B_\delta''$ can be expanded as a sum over Kraus operators, $\{\chi_\delta^k\},$ acting on the density matrix. 

Next, $\chi^k_\delta$ can be expanded as a sum over 4-qubit Pauli operators (including the identity), $\{\sigma_i\}$, weighted by complex coefficients, so that $\chi^k_\delta = \sum w_i^k \sigma_i$, with $w_i^k \in \mathbb{C}$ and $\sum_k \sum_i w_i^k w_i^{k*} = 1$. Table \ref{tab:stabscom} shows whether a given Pauli term in the deviation operator commutes or anticommutes with each trap setting, and hence whether such an error is detectable or not. We note that the only Pauli terms which commute with the measurement and terms which correspond to a simultaneous bit-flip error on only the first and last qubits remain undetected. The first set of terms leave the computation unaltered, and hence do not represent errors. However the latter group do represent an error which cannot be detected by our current setup.

\begin{table}[h]
	\centering
\begin{tabular}{|c|c|c|c|c|}
\hline
Pauli ($\sigma_i$) &  \multicolumn{3}{|c|}{Trap Stabilizer Measurement} & Overall\\
&$X \otimes \mathbb{I} \otimes Y \otimes Y$& $Y \otimes X \otimes X \otimes Y$ & $Y \otimes Y \otimes \mathbb{I} \otimes X$ &\\
\hline
$C \otimes C \otimes C \otimes C$ & \tick & \tick & \tick & \tick\\
$C \otimes C \otimes C \otimes A$ & \cross & \cross & \cross & \cross \\
$C \otimes C \otimes A \otimes C$ & \cross & \cross & \tick & \cross\\
$C \otimes C \otimes A \otimes A$ & \tick & \tick & \cross & \cross\\
$C \otimes A \otimes C \otimes C$ & \tick & \cross & \cross & \cross \\
$C \otimes A \otimes C \otimes A$ & \cross & \tick & \tick & \cross\\
$C \otimes A \otimes A \otimes C$ & \cross & \tick & \cross & \cross\\
$C \otimes A \otimes A \otimes A$ & \tick & \cross & \tick & \cross\\
$A \otimes C \otimes C \otimes C$ & \cross & \cross & \cross & \cross\\
$A \otimes C \otimes C \otimes A$ & \tick & \tick & \tick & \tick\\
$A \otimes C \otimes A \otimes C$ & \tick & \tick & \cross & \cross\\
$A \otimes C \otimes A \otimes A$ & \cross & \cross & \tick & \cross\\
$A \otimes A \otimes C \otimes C$ & \cross & \tick & \tick & \cross\\
$A \otimes A \otimes C \otimes A$ & \tick & \cross & \cross & \cross\\
$A \otimes A \otimes A \otimes C$ & \tick & \cross & \tick & \cross\\
$A \otimes A \otimes A \otimes A$ & \cross & \tick & \cross & \cross\\
\hline
\end{tabular}
\caption{Pauli terms in the deviation operator $B''_\delta$ and whether or not they are detected by a particular trap setup or not. Although there are 256 distinct 4-qubit Pauli operators, including the identity, these can be grouped into 16 distinct sets based on whether each local term commutes ($C \in \{\mathbb{I},Z\}$) or anticommutes ($A \in \{X,Y\}$) with the computational basis measurement carried out immediately after the deviation operator acts. Note that all such terms are either leave the computation invariant, or are detected by at least one trap setting, with the exception of $A\otimes C\otimes C \otimes A$. \label{tab:stabscom}}
\end{table}

While this appears to be an insurmountable problem if we wish to verify a general quantum computation, the problem disappears entirely if we consider only those computations for which the output of the computation only depends on the parity of the measurement results of qubits 1 and 4. This is because flipping both measurement outcomes leaves their parity invariant, and hence the outcome of the computation remains the same. Thus for the remainder of this section we consider only those computations for which simultaneously flipping the first and last measurement results leave the outcome of the computation invariant.

With this restriction in place, we take the verification protocol to proceed as follows. First the verifier randomly chooses whether or not to perform a computation as normal or instead to perform a trap computation. We assume that a trap computation is chosen with probability $p$. Next the verifier chooses uniformly at random an index for the trap qubit.
As traps 2 and 3 correspond to the same stabilizer measurement we would obtain a better probability of detecting an error by choosing between the three stabilizer measurements uniformly at random. However, here we use an identical probability for choosing each trap index, since this is optimal in the case where our experimental restrictions are limited and we can employ the full protocol of~\cite{Fitzsimons2012}.

We wish to bound the probability that a given run of the computation yields the correct results based on the probability of trap computations yielding incorrect results. To do this, we note that for the set of computations we consider, any Pauli term in $B''_\delta$ which leads to an error in the outcome of the computation necessarily anticommutes with at least one of the trap stabilizer measurements and hence is detected with probability at least $p/4$. Thus any deviation which flips at least one of the measurement outcomes is detected with probability at least $p/4$. If the probability that a malicious prover flips one or more measurement outcomes is $\epsilon$, then the probability that a trap computation yields the correct result is $\langle t \rangle \geq \epsilon p / 4$. Thus the probability that the outcome of a computation is incorrect is bounded from above by $\epsilon \leq \frac{4\langle t \rangle}{p}$.

We note that in order for the above verification procedure to work, it is necessary for all qubits to be fully blind (i.e. all possible choices of $\theta$ and $\delta$ from $\{0, \frac{\pi}{4}, \frac{\pi}{2}, \frac{3\pi}{4}, \pi, \frac{5\pi}{4}, \frac{3\pi}{2}, \frac{7\pi}{4}\}$ should be possible for each qubit). In the current generation of experiments this property holds only for qubits 2 and 3, and the value for $\delta_1$ and $\delta_4$ are fixed. However we note that these fixed values do represent a legitimate choice on the part of the verifier, and as long as the prover does not have a priori information about this restriction the proof of authentication holds.


\subsection{Experimental settings}
In our experiment, we choose the set of phases and measurement setting as given in Table~\ref{table:trapqubits} to prepare traps on all qubits:
\begin{table}[h!]
	\centering
\begin{tabular}{lll}
\hline
\multirow{2}{*}{$\ket{\mbox{trap}_1}=\ket{-_i}$:} 
& $\theta_1=0$, $\theta_2=\pi/2$, $\theta_3=0$, $\theta_4=0$  \\ 
& $\delta_1=\delta_{\mbox{trap}}$, $\delta_2=-\pi/2$, $\delta_3=\pi$, $\delta_4=-\pi/2$\\\hline
\multirow{2}{*}{$\ket{\mbox{trap}_1}=\ket{+}$:} 
& $\theta_1=0$, $\theta_2=\pi/2$, $\theta_3=3\pi/2$, $\theta_4=0$  \\ 
& $\delta_1=\delta_{\mbox{trap}}$, $\delta_2=-\pi/2$, $\delta_3=5\pi/4$, $\delta_4=\pi/2$\\\hline
\multirow{2}{*}{$\ket{\mbox{trap}_2}=\ket{+}$:}
& $\theta_1=0$, $\theta_2=\pi/2$, $\theta_3=\pi$, $\theta_4=0$ \\
& $\delta_1=-\pi/2$, $\delta_2=\delta_{\mbox{trap}}$, $\delta_3=0$, $\delta_4=0$\\\hline
\multirow{2}{*}{$\ket{\mbox{trap}_2}=\ket{+_i}$:}
& $\theta_1=0$, $\theta_2=0$, $\theta_3=0$, $\theta_4=0$ \\
& $\delta_1=\pi/2$, $\delta_2=\delta_{\mbox{trap}}$, $\delta_3=0$, $\delta_4=0$\\\hline
\multirow{2}{*}{$\ket{\mbox{trap}_3}=R_z(3\pi/4)\ket{+}$:} 
& $\theta_1=0$, $\theta_2=\pi/2$, $\theta_3=5\pi/4$ , $\theta_4=0$ \\
& $\delta_1=\pi$, $\delta_2=-\pi/2$, $\delta_3=\delta_{\mbox{trap}}$, $\delta_4=\pi/2$\\\hline
\multirow{2}{*}{$\ket{\mbox{trap}_3}=R_z(\pi/4)\ket{+}$:} 
& $\theta_1=0$, $\theta_2=\pi/2$, $\theta_3=7\pi/4$ , $\theta_4=0$ \\
& $\delta_1=\pi$, $\delta_2=0$, $\delta_3=\delta_{\mbox{trap}}$, $\delta_4=-\pi/2$\\\hline
\multirow{2}{*}{$\ket{\mbox{trap}_4}=\ket{-}$:}
& $\theta_1=0$, $\theta_2=\pi/2$, $\theta_3=\pi/4$, $\theta_4=0$ \\
& $\delta_1=\pi/2$, $\delta_2=0$, $\delta_3=5\pi/4$, $\delta_4=\delta_{\mbox{trap}}$\\\hline
\multirow{2}{*}{$\ket{\mbox{trap}_4}=\ket{-_i}$:}
& $\theta_1=0$, $\theta_2=\pi/2$, $\theta_3=3\pi/4$, $\theta_4=0$ \\
& $\delta_1=\pi/2$, $\delta_2=\pi/2$, $\delta_3=7\pi/4$, $\delta_4=\delta_{\mbox{trap}}$\\\hline
\end{tabular}
\caption{\label{table:trapqubits} Blind phases and measurement instructions for the entanglement verification procedure.}
\end{table}

\section{Entanglement verification}

As discussed in the main paper we demonstrate how our restricted verification scheme can be exploited for the verification of a non-classical computation, in the form of a measurement of Bell statistics. For a test of Bell's inequality, the certain measurements $\alpha, \alpha'$ and $\beta, \beta'$ need to be performed on a two-qubit state $\ket{\psi}_{a,b}$, where $\alpha, \alpha'$ ($\beta, \beta'$) are the measurements performed on qubit $a$ ($b$).
If the state $\ket{\psi}_{a,b}$ is entangled, a maximal violation of the Bell inequality of the Clauser-Horne-Shimony-Holt  (CHSH)-type,
\begin{equation}
	S=|E(\alpha, \beta)-E(\alpha, \beta')|+|E(\alpha', \beta)+E(\alpha', \beta')|\leq2,
\end{equation}
can be obtained.
Here the correlation coefficients are defined as
\begin{equation}
E(\alpha, \beta)=\frac{C_{00}(\alpha, \beta)-C_{01}(\alpha, \beta)-C_{10}(\alpha, \beta)+C_{11}(\alpha, \beta)}{C_{00}(\alpha, \beta)+C_{01}(\alpha, \beta)+C_{10}(\alpha, \beta)+C_{11}(\alpha, \beta)}
\end{equation}
and $C_{ij}(\alpha, \beta)$ are the coincidence counts for obtaining measurement results $i=\{0,1\}$ on qubit $a$ and $j=\{0,1\}$ on qubit $b$ for measurements in bases $\alpha$ and $\beta$ on qubits $a$ and $b$ respectively.


%

\begin{figure}
\centering
    \includegraphics[width=0.2\textwidth]{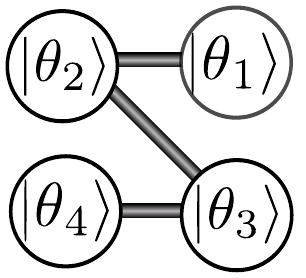}
  \caption{Blind zigzag cluster}
  \label{fig:cluster}
\end{figure}

We exploit the framework of blind quantum computing, in order to enable a verifier to perform a blind Bell test.
We choose the blind cluster state to be a zigzag cluster state, shown in Figure~\ref{fig:cluster}.

The underlying circuit, which is obtained, when measurements in the basis $\ket{\pm_{\delta_j}}=(\ket{0}\pm e^{i\delta_j}\ket{1})/\sqrt{2}$ are performed on the blind zigzag cluster state with blind qubits being in the state $\ket{\theta_j}$ is given by the circuit below
\begin{equation}
	\Qcircuit @C=1.0em @R=.7em {
	\ket{+}&&\qw                   &\qw       &\ctrl{1} &\qw  &\gate{R_z(-\delta_2+\theta_2)}&\gate{H} &\gate{R_z(\theta_1)}  &\measureD{\delta_1} \\
	\ket{+}&&\gate{R_z(-\delta_4+\theta_4)}&\gate{H}  &\ctrl{0} &\qw &\gate{R_z(\theta_3)}&\qw   &\qw         &\measureD{\delta_3}\\
	&&\mbox{\raisebox{-1.6em}{\textit{State generation}}}&&&&&\mbox{\raisebox{-1.6em}{\textit{Bell test}}}
	\gategroup{1}{1}{2}{5}{2em}{--}\gategroup{1}{7}{2}{10}{.7em}{--},
	}
\end{equation}

\vspace{0.3cm}
\noindent where
\raisebox{0.4em}{$
\Qcircuit @C=0.9em @R=.6em {
                      &\ctrl{1}&\qw    \\
                       &\ctrl{0}&\qw    
}
$}
denotes a CPhase gate ($\mbox{CPhase}\ket{ij}=(-1)^{ij}\ket{ij}$) and 
$
\Qcircuit @C=1.0em @R=.6em {
&\qw   &\measureD{\phi}
}
$
a measurement in the basis $\ket{+_{\phi}}$. Here, $R_z(\phi) =  \exp(-i\phi\sigma_z/2)$,  $H=(\sigma_x + \sigma_z)/\sqrt{2}$ and $\sigma_x, \sigma_y$ and $\sigma_z$ denote the usual Pauli matrices.

As we will see in the following, a blind Bell test can be implemented by choosing suitable combinations of $\delta_j$ and $\theta_j$. For this, the left part of the circuit (shown above) implements the state generation, whereas the right parts realizes the Bell test.


The verifier can choose between entangled states and product states for the Bell test.
For example, by choosing $-\delta_4+\theta_4=0$ (or $\pi$), the input state will be a product state:

\begin{equation}
	\Qcircuit @C=1.0em @R=.7em {
	\ket{+}&&\qw                   &\qw       &\ctrl{1} &\qw&\rstick{\cdots} &&&{\raisebox{-2em}{$\Rightarrow$}}&&&& \ket{+}&    &\ctrl{1} &\qw&\rstick{\cdots}\\
	\ket{+}&&\gate{R_z(0)}&\gate{H}  &\ctrl{0} &\qw&\rstick{\cdots}&&&&&&&\ket{0}&  &\ctrl{0} &\qw&\rstick{\cdots}\\
	}
\end{equation}
Alternatively, by choosing $-\delta_4+\theta_4=\pi/2$ (or $-\pi/2$), the input state will be an entangled state:
\begin{equation}
\Qcircuit @C=1.0em @R=.7em {
\ket{+}&&\qw                   &\qw       &\ctrl{1} &\qw&\rstick{\cdots} &&&{\raisebox{-2em}{$\Rightarrow$}}&&&&\ket{+}&    &\ctrl{1} &\qw&\rstick{\cdots}\\
\ket{+}&&\gate{R_z(\pi/2)}&\gate{H}  &\ctrl{0}&\qw&\rstick{\cdots}&&&&&&&\ket{-_i}&  &\ctrl{0} &\qw&\rstick{\cdots}\\
}
\end{equation}

\noindent This demonstrates that the verifier can choose between different---product or entangled---input states for the Bell test by choosing different combinations of $\delta_4$ and $\theta_4$. For our demonstration, we exemplarily choose the entangled state to be $\mbox{CPhase}\ket{+}\ket{-_i}$.

In the blind framework, the Bell measurements are determined by the choice of the blind phases $\theta_1$, $\theta_2$, and $\theta_3$ as well as by the measurement settings $\delta_1$, $\delta_2$, and $\delta_3$ on the blind zigzag cluster state.
For our demonstration, we choose the Bell measurement angles as given in the main paper.

\vspace{0.3cm}
The Bell settings of $\alpha$ and $\alpha'$ are determined by the choice of $(-\delta_2+\theta_2)$ and
$(\delta_1-\theta_1)$. For the Bell measurement angle $\alpha=\pi/2$, we choose $\delta_1-\theta_1=\pi/2$ and $-\delta_2+\theta_2=\pi$.
\begin{equation}
	\Qcircuit @C=1.0em @R=.7em {
	\ket{+}&    &\ctrl{1} &\gate{R_z(\pi)}&\gate{H}  &\measureD{\pi/2}&&&{\raisebox{-2em}{$\Rightarrow$}}&&&& \ket{+}&    &\ctrl{1} &\gate{R_z(\pi)}   &\measureD{-\pi/2}\\
	\ket{-_i}&  &\ctrl{0}  &\qw&\rstick{\cdots}  &&&&&&&& \ket{-_i}&  &\ctrl{0} &\qw &\rstick{\cdots},\\
	}
\end{equation}
which leads to a measurement in the basis $\alpha=\pi/2$ in the upper wire:
\begin{equation}
\Qcircuit @C=1.0em @R=.7em {
{\raisebox{-2em}{$\Rightarrow$}}&&&\ket{+}&    &\ctrl{1} &\qw &\qw  &\measureD{\pi/2} \\
&&&\ket{-_i}&  &\ctrl{0} &\qw &\rstick{\cdots}\\
}
\end{equation}

\vspace{0.3cm}
For the Bell measurement angle $\alpha'=\sigma_z$, we choose $\delta_1-\theta_1=0$. 
With that configuration, $-\delta_2+\theta_2$ can have any value, since a $R_z(-\delta_2+\theta_2)$ rotation does not affect the state $\ket{0}$:
\begin{equation}
	\Qcircuit @C=1.0em @R=.7em {
	\ket{+}&    &\ctrl{1} &\gate{R_z(-\delta_2+\theta_2)}&\gate{H}  &\measureD{0} &&&{\raisebox{-2em}{$\Rightarrow$}}&&&&\ket{+}&    &\ctrl{1} &\gate{R_z(-\delta_2+\theta_2)} &\measureD{\sigma_z}\\
	\ket{-_i}&  &\ctrl{0}  &\qw&\rstick{\cdots}  &&&&&&&&\ket{-_i}&  &\ctrl{0}&\qw  &\rstick{\cdots} \\
	}
\end{equation} 
Finally, we obtain a measurement in the basis $\alpha'=\sigma_z$:
\begin{equation}
\Qcircuit @C=1.0em @R=.7em {
{\raisebox{-2em}{$\Rightarrow$}}&&&\ket{+}&    &\ctrl{1} &\qw &\qw  &\measureD{\sigma_z} \\
&&&\ket{-_i}&  &\ctrl{0}&\qw &\rstick{\cdots}\\
}
\end{equation}

\vspace{0.3cm}
The angles $\beta$ and $\beta'$ are determined by $\delta_3$ and $\theta_3$.
\begin{equation}
\Qcircuit @C=1.0em @R=.7em {
\ket{+}&    &\ctrl{1} &\qw &\rstick{\cdots}&&&&{\raisebox{-2em}{$\Rightarrow$}}&&& \ket{+}&    &\ctrl{1} &\qw &\rstick{\cdots} \\
\ket{-_i}&  &\ctrl{0} &\gate{R_z(\theta_3)}&\qw   &\qw          &\measureD{\delta_3}&&&&&\ket{-_i}&  &\ctrl{0} &\qw&\qw           &\measureD{\delta_3-\theta_3}\\
}
\end{equation}
To choose the Bell settings, $\beta=-3\pi/4$ and $\beta'=-\pi/4$, we simply take $\delta_3-\theta_3$ to be equal to $\beta$ or $\beta'$.

\subsection{Experimental measurement settings}
In our experiment, we choose the settings given in table~\ref{table:belltest}.
\begin{table}[h!]
	\centering
\begin{tabular}{lll}
\hline
\multirow{2}{*}{$\alpha$, $\beta$:} & $\theta_1=0$, $\theta_2=\pi/2$, $\theta_3=3\pi/4$, $\theta_4=0$  \\
																		& $\delta_1=\pi/2$, $\delta_2=-\pi/2$, $\delta_3=0$, $\delta_4=-\pi/2$\\\hline
\multirow{2}{*}{$\alpha$, $\beta'$:}& $\theta_1=0$, $\theta_2=0$, $\theta_3=3\pi/4$, $\theta_4=0$ \\
																		& $\delta_1=\pi/2$, $\delta_2=0$, $\delta_3=-\pi/2$, $\delta_4=-\pi/2$\\\hline
\multirow{2}{*}{$\alpha'$,$\beta$:}&  $\theta_1=0$, $\theta_2=\pi/2$, $\theta_3=\pi/4$ , $\theta_4=0$ \\
																		& $\delta_1=0$, $\delta_2=-\pi/2$, $\delta_3=-\pi/2$, $\delta_4=-\pi/2$\\\hline
\multirow{2}{*}{$\alpha'$, $\beta'$:}& $\theta_1=0$, $\theta_2=0$, $\theta_3=\pi/4$, $\theta_4=0$ \\
																			& $\delta_1=0$, $\delta_2=0$, $\delta_3=0$, $\delta_4=-\pi/2$\\\hline
\end{tabular}
\caption{Blind phases and measurement instructions for the preparation of a set of trap qubits}
\label{table:belltest}
\end{table}
	

Note, that for the second setting $\alpha$, $\beta'$ we measure  $\alpha+\pi$ and $\beta'+\pi$ instead of $\alpha$ and $\beta'$. 
This has no effect on the Bell inequality since only the measurement outcomes are exchanged
($00\rightarrow11$, $01\rightarrow10$, $10\rightarrow01$, $11\rightarrow00$).
This exchange of the measurements outcomes can be interpreted as the verifier choosing $r_j=1$.
In the blind quantum computing framework, $r_j$ is a randomly chosen value in $\{0,1\}$ which hides the value of the measurement
outcome.

\subsection{Bell test verification}

In order to show that the Bell test is invariant under errors of the form $A\otimes C \otimes C \otimes A$, as required to show verification in our setting, we note that the circuit implemented by our measurement settings is described by the circuit below.

\[
	\Qcircuit @C=0.2em @R=.7em {
	\lstick{\ket{+}}&&\qw                   &\qw    &\qw    &\ctrl{1}&\gate{R_z(-\delta_2+\theta_2)} &\gate{Z^{m_1}} &\gate{H} &\gate{R_z(\theta_1)} &\gate{Z^{m_1}} & \measureD{\delta_1} \\
	\lstick{\ket{+}}&&\gate{R_z(-\delta_4+\theta_4)} &\gate{Z^{m_4}} &\gate{H}  &\ctrl{0} &\gate{R_z(\theta_3)}&\gate{Z^{m_3}} &\measureD{\delta_3}
	}
\]

Note that an error of this form flips both $m_1$ and $m_4$. The effect of flipping $m_1$ is trivially identical to flipping the outcome of the first logical qubit in the Bell test. Although it is not immediately obvious, we note that since $Z^{m_1} R_Z(-\delta_4 + \theta_4)= R_Z(\pm \frac{\pi}{2})$ and $H Z R_Z(\pm \frac{\pi}{2}) \ket{+} = Z H R_Z(\pm \frac{\pi}{2}) \ket{+}$, a bit flip error on $m_4$ leads to a bit flip error in the outcome of the measurement result for the second logical qubit. Thus all errors of the form $A\otimes C \otimes C \otimes A$ flip the outcome of both measurements in a Bell test. However, we note that the outcome of the Bell test depends only on the parity of these two measurements, and hence any inferred value of the CHSH quantity is left unchanged by such errors.

\end{widetext}
\end{document}